\documentclass{aa}
\usepackage{psfig}
\usepackage{graphicx}
\begin{document}

   \title{Symbiotic stars on Asiago archive plates.  II}

   \author{
           Ulisse Munari\inst{1}
	   \and
	   Rajka Jurdana-\v{S}epi\'c\inst{2}
          }

   \offprints{U.Munari \\ \email{munari@pd.astro.it}}

   \institute{
Osservatorio Astronomico di Padova, Sede di Asiago,
I-36012 Asiago (VI), Italy
\and
Physics Department, University of Rijeka, Omladinska 14, HR-51000 Rijeka,
Croatia
             }

   \date{Received 20 November 2001 / Accepted 22 January 2002}

   \abstract{The Asiago photographic archive has been searched for plates
containing the symbiotic stars Hen~2-468, QW~Sge, LT~Del, V407~Cyg, K~3-9,
V335~Vul, FG~Ser and Draco~C-1. A total of 635 plates imaging the program
stars have been found and the brightness estimated using the Henden \&
Munari (2000) {\sl UBV(RI)$_{\rm C}$} photometric sequences. These historical 
data have allowed for the first time the determination of the orbital periods of
Hen~2-468 (774 days) and QW~Sge (390.5 days), a significant improvement in
the orbital period of LT~Del (465.6 days) and for V407~Cyg an evaluation of
the Mira's pulsation period and complex lightcurve shape in the
red ($R$ and $I$ bands). Some previously unknown outbursts have been discovered too.
   \keywords{binaries:symbiotic}
   }

   \maketitle

\section{Introduction}

In Paper~I (Munari et al. 2001) we presented the first results of the search
for plates in the Asiago photographic archive containing symbiotic stars. A
total of 602 plates imaging AS~323, Ap~3-1, CM~Aql, V1413~Aql, V443~Her,
V627~Cas and V919~Sgr were located and the brightness of symbiotic stars
estimated at the microscope against the {\sl UBV(RI)$_{\rm C}$} comparison
sequences calibrated by Henden \& Munari (2000). The effort paid fruitful
dividends on these poorly known systems: AS~323 was discovered to be quite
probably an eclipsing system with the shortest known orbital period among
symbiotic stars (P=197.6 days), and CM~Aql was found to have a P=1058 days
orbital period and a remarkable amplitude of the associated
reflection/heating effect ($\Delta B \sim 2.2$ mag).

In this second paper of the series we searched the Asiago archive for plates imaging
the eight symbiotic stars Hen~2-468, QW~Sge, LT~Del, V407~Cyg, K~3-9,
V335~Vul, FG~Ser and Draco~C-1. A total of 635 plates were found containing the
program stars and their brightness has been again estimated at the
microscope against the {\sl UBV(RI)$_{\rm C}$} comparison sequences
calibrated by Henden \& Munari (2000).

The time scale of variability for symbiotic stars is quite long: the orbital
periods range from $\sim$1 year up to several decades while rise and decay
from an outburst may take anything from a few years to more than a century
(cf. Kenyon 1986, Mikolajewska 1996). Such times scales are generally too
long for devoted photometric programs to be carried out by a single
Institute, but are instead well handled by those patrol programs that
run for at least several decades at some Observatories around the world, one
of them being Asiago. To stimulate and facilitate inspection of plates at
the various archives in the world, Henden \& Munari (2000, 2001, 2002) have
so far calibrated accurate and deep {\sl UBV(RI)$_{\rm C}$} sequences around
60 symbiotic stars. The present series of papers applies these sequences to
those plates in the Asiago archive found to contain symbiotic stars, with the
aim of contributing our data to the effort of reconstructing the
longest and most detailed possible historical lightcurves.

\section{Results}

\subsection{Hen 2-468}

Discovered on objective-prism plates by Henize (1967) who noted a possible
variability, its symbiotic nature was recognized by Carrasco et al.
(1983) and Allen (1984) on the basis of optical spectra showing a late
M-giant continuum and a rich emission line spectrum including HeII and the
6830~\AA\ symbiotic band. Allen (1974) determined its infrared brightness as
$K=7.96$, $J-K=+1.48$. It was discovered to be a variable star by Margoni and
Stagni (1984, their variable \#3) that reported for the period 1969--1979 a
brightness range 14.0 $\leq V \leq$15.0 and 16.0$\leq B \leq$17.2. The star
however has not yet received a variable star name. Infrared photometry by
Munari et al. (1992) gave $K=8.02$, $J-K=+1.50$, almost identical to Allen
(1974), suggesting that the cool giant in He2-468 is not intrinsically
variable. Henden and Munari (2000) observed Hen~2-468 in quiescence at
$B$=16.6 and $B-V$=+1.8 in 1999, in agreement with the Margoni and Stagni (1984)
values.

\begin{figure}
\centering
\includegraphics[width=8.7cm]{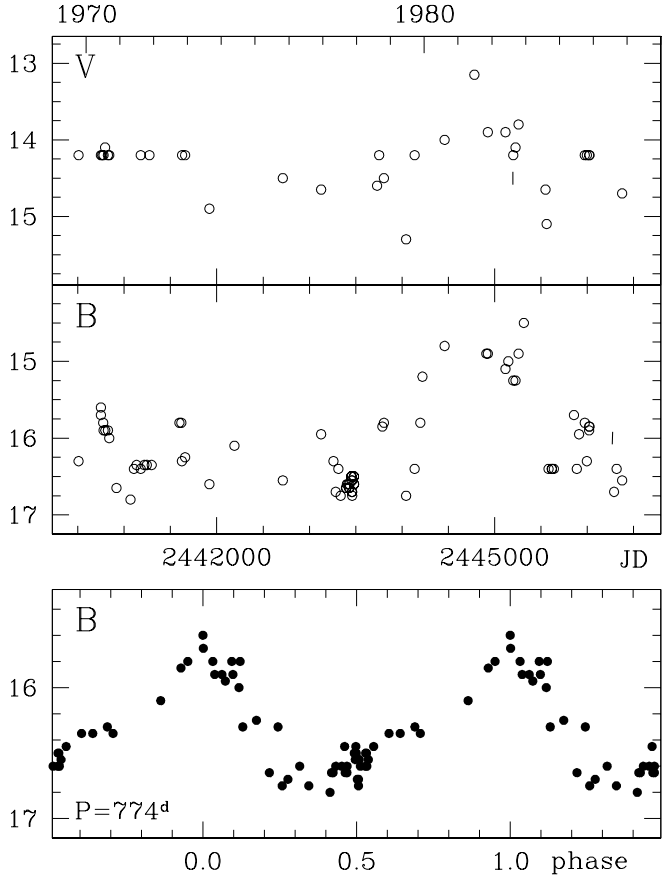}
      \caption{$B$ and $V$ lightcurve for Hen~2-468. Vertical bars indicate 
               ``fainter than".{\sl Lower panel}: phase plot of $B$ quiescence 
                 data (between 2440508 and 2444044) according to Eq. (1).}
\label{He2_468}
\end{figure}

\begin{table}
\centering
\caption[]{The program symbiotic stars. Equinox J2000 and epoch $\sim$2000
coordinates are from Henden \& Munari (2000).}
\begin{tabular}{llr}
\hline
          & \multicolumn{1}{c}{$\alpha_{J2000}$} & 
            \multicolumn{1}{c}{$\delta_{J2000}$} \\  
          &               &              \\
Draco C-1 & 17 19 57.661  & +57 50 05.74 \\
FG Ser    & 18 15 07.121  &--00 18 51.96 \\
K 3-9     & 18 40 24.133  &--08 43 57.73 \\
V335 Vul  & 19 23 14.124  & +24 27 40.17 \\
QW Sge    & 19 45 49.548  & +18 36 48.47 \\
LT Del    & 20 35 57.234  & +20 11 27.91 \\
Hen 2-468 & 20 41 18.989  & +34 44 52.52 \\
V407 Cyg  & 21 02 09.831  & +45 46 32.85 \\
\hline
\end{tabular}
\label{tab1}
\end{table}

The Asiago plates (78 $B$, 34 $V$ and 5 $I$ band) cover the time span
between Oct. 1969 and Dec. 1987. The resulting $B$ and $V$ lightcurves are
plotted in Figure~1. On them, Hen~2-468 appears to have been in quiescence up
to Oct. 1979, when it entered an outburst phase that lasted until the summer
of 1985 when the system returned to mean quiescence brightness. The outburst
amplitude was limited, at maximum reaching $\Delta B \sim$1.8 mag
above mean quiescence level. Given the red color in quiescence ($B-V$=+1.8,
cf. Table~2), the outburst is barely traceable in the $V$ lightcurve
dominated by the stable emission of the M giant ($B-V \sim$+1.0 at outburst
maximum). This is the first-ever recorded outburst of Hen~2-468.

The $B$ quiescence lightcurve of Hen~2-468 looks stable in mean brightness
($< B >$=16.3) but highly variable around it. A period search was performed
and revealed a strong 774 day periodicity. The absence of a corresponding
modulation in the $V$ data and the shape of the phased quiescence $B$
lightcurve (presented in the bottom panel of Figure~1) both support the
interpretation of the $\Delta B \sim$1.1 mag variability as a
reflection/heating effect, with
\begin{equation}
Max(B)\ =\ 2440749 +\ 774\times E
\end{equation}
giving the times of maxima, corresponding to the passage at inferior
conjunction of the white dwarf companion to the M giant. The asymmetric
placement of the minimum at $\phi\sim$0.35 could argue in favor of a
moderately eccentric orbit. The present one is the first ever determination
of the orbital period for Hen~2-468.

\begin{figure}
\centering
\includegraphics[width=8.7cm]{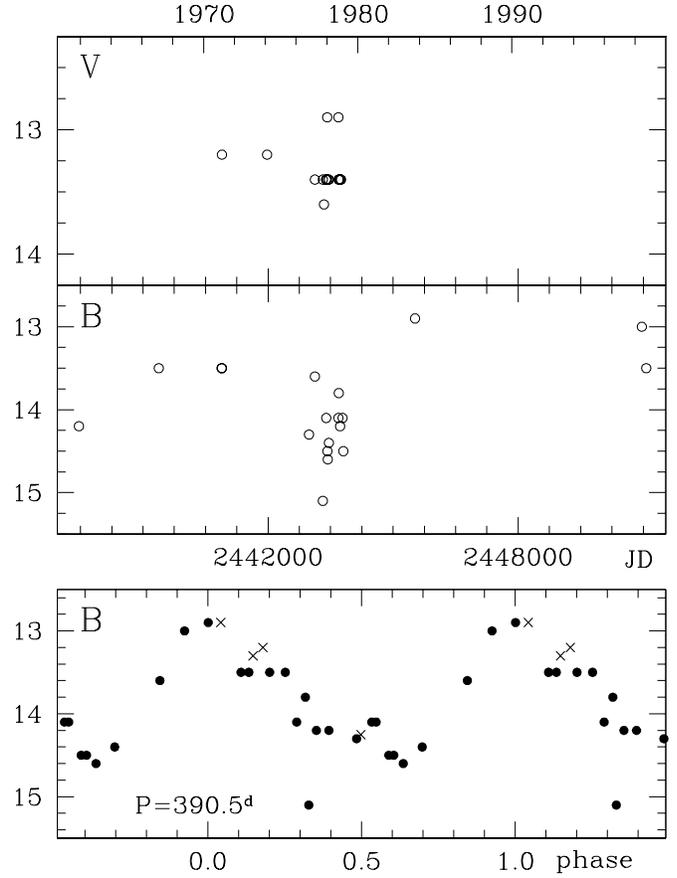}
      \caption{$B$ and $V$ data for QW Sge. {\sl Lower panel}: phase
      plot of $B$ data according to Eq. (2). Crosses indicate data from Henden
      \& Munari (2000), Munari \& Buson (1991), Palomar DSS-1 and DSS-2.}
\label{QW Sge}
\end{figure}

\subsection{QW~Sge}

QW~Sge (= AS~360) is another symbiotic star that has received scanty
attention in the literature. Munari \& Buson (1991) already reviewed the
latter and discussed the system properties based on IUE, optical and IR
observations.

QW Sge has an optical companion 3.5 arcsec to the north, that Munari and
Buson classified as an F0~V star with $B$=13.59 and $B-V$=+0.45. Henden and
Munari (2000) found different values for it ($B$=13.18 and {\sl
B--V}=+0.83), with a large scatter of 0.25 mag between three different
measurements (compared to the stability at a few millimag for nearby stars
of similar brightness). This clearly indicates that the optical companion is
itself a variable star, and this complicates the interpretation of
photometry made with moderate or short focus telescopes that are not able to
separate QW~Sge from the close optical companion (as it is the case for most
of the photographic plates in the archives around the world).

The 464 blue plates from the archive of the Sternberg Astronomical
Institute in Moscow examined by Kurochkin (1993) suffer from this blending.
All his measurements refer to QW~Sge and the companion star merged in a
single stellar image. Kurochkin's lightcurve covers the period 1898--1990,
during which two outbursts took place: one extending from July 1962 to March
1972 with $B$=11.5 at maximum, the other from May 1982 to September 1989
with a much more complex lightcurve and a peak brightness $B$=12.0.
In between the combined brightness of QW~Sge and its nearby companion remains
in the range 12.9$\leq B \leq$13.4

\begin{figure}
\centering
\includegraphics[width=8.8cm]{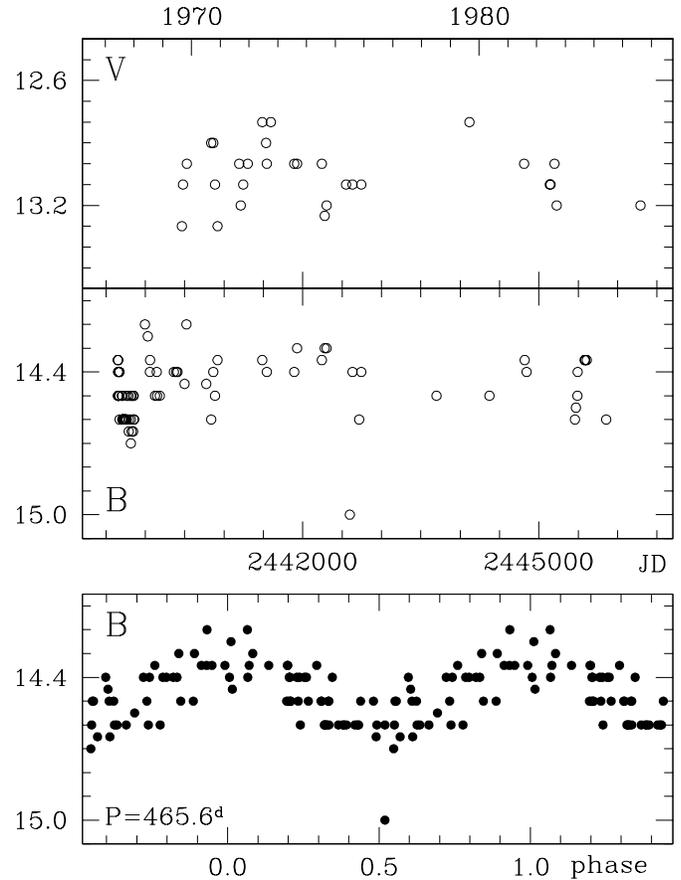}
      \caption{$B$ and $V$ lightcurve of LT Del. {\sl Lower panel}: phase 
               plot of $B$ data according to Eq. (3).}
\label{LTDel}
\end{figure}

Our measurements in Table~2 pertain to quiescence conditions for QW~Sge. The
favorable focal lengths of the Asiago Schmidt telescopes allow a marginal
separation of QW~Sge from the companion star. The measurements in Table~2
pertain to QW~Sge alone. A few more plates from the Asiago archive were not
considered here because bad seeing conditions caused an unrecoverable
merging of the two stellar images into a single one.

A period search was performed and a fine periodicity emerged following 
the ephemeris
\begin{equation}
Max(B)\ =\ 2445528 +\ 390.5\times E
\end{equation}
The Table~2 data are phase plotted in Figure~2. Our measurements are limited
in number and non-negligible errors are expected to affect our estimates of
the nearly merged stellar pair. To test ephemeris (2) we therefore
estimated QW~Sge brightness on Palomar DSS-1 and DSS-2 blue plates
(JD=2433480 and $B$=13.3, JD=2446326 and $B$=12.9, respectively) and plotted
them in Figure~2 as crosses together with single $B$ measurements from
Henden and Munari (2000) and Munari and Buson (1991). The agreement of these
external data with those from the Asiago plates is excellent over the 5 decades
spanned by the data. The lightcurve shape, its amplitude ($\Delta B$=1.7
mag, and a few tenths in $V$) and the high repeatability over a long period
of time suggests an interpretation in term of reflection/heating effect in
phase with the orbital motion. The present 390$_{.}^{d}$5 day is the first
ever determination of the orbital period of QW~Sge.

\begin{figure}
\centering
\includegraphics[width=8.8cm]{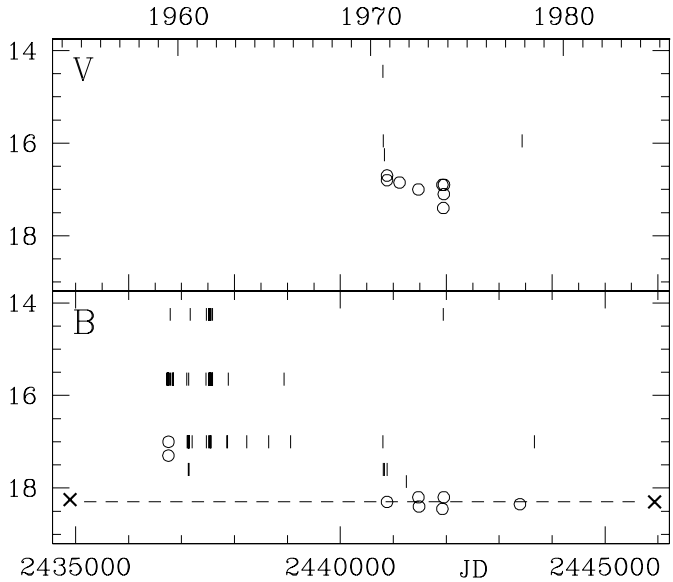}
      \caption{$B$ and $V$ data of K~3-9. Vertical bars indicate
               {\sl fainter than}. The two crosses indicate the brightness
               of K~3-9 on Palomar DSS-1 and DSS-2 blue plates, and the 
               dashed line the 1999 quiescence $B$=18.3 level.}
\label{K39}
\end{figure}

\subsection{LT Del}

The history and photometric properties of LT~Del up to the early 90ies have
already been reviewed by Munari and Buson (1992). The quiescent lightcurve
of LT~Del displays a strong reflection/heating effect with amplitudes $\Delta
U$=+1.7, $\Delta B$=0.5 and $\Delta V$=0.2 mag, that Arkhipova and Noskova
(1988) reported to follow the ephemeris {\sl Min = 2445910 + 488$\times$E},
later revised to {\sl Min = 2445910 + 478.2$\times$E} by Arkhipova et al.
(1995). The latter listed $B$=14.4 and $B-V$=+1.3 as mean values for
quiescence. Passuello et al. (1994) discovered the only outburst so far
recorded for LT~Del, with a $\Delta B\sim$1.6 mag in 1994.

Our lightcurve of LT~Del extends from Sept. 1967 to Aug. 1985 (cf.
Figure~2), with mean values $B$=14.41 and $B-V$=+1.32 identical to those of
Arkhipova et al.  (1995) for LT~Del in quiescence. A period search revealed
a strong periodicity at 465$_{.}^{d}$6, with an amplitude $\Delta B
\sim$0.27 mag and a sinusoidal shape indicative of a reflection/heating
effect. The variation follows the ephemeris
\begin{equation}
Max(B)\ =\ 2440493 +\ 465.6\times E
\end{equation}
The 465$_{.}^{d}$6 period differs significantly from previous estimates. The
488 days suggested by Arkhipova and Noskova (1988) is definitively ruled out
by our data, while the 478.2 days indicated by Arkhipova et al. (1995)
performs only marginally better with a high dispersion affecting the folded
lightcurve.

\subsection{K 3-9}

The photometric properties, history and orbital period are unknown for this
faint and poorly studied symbiotic star. Given its optical faintness, the
scanty investigations were concentrated in the infrared and radio domains
where the system is much easier to observe (Ivison and Seaquist 1995).
Henden \& Munari (2000) report an average $B$=18.3 and $B-V$=+1.3 for
1999. 

The lightcurve from Asiago archival plates presented in Table~2 and Figure~4
covers the period from 1959 to 1979 and confirms the optical faintness and
the moderately red color of K~3-9. The object has always been close to plate
limiting magnitude, however a flat quiescence in the 70ies with $B$=18.3 and
$B-V \sim$1.0 and a brighter phase in 1959 at $B$=17.2 are well established.
A $\sim$0.5 mag variability affected K~3-9 in $V$ with apparently no
counterpart in $B$.

Ivison and Seaquist (1995) argued about the possible presence of a Mira in
K~3-9 and an ongoing symbiotic nova outburst phase for the WD companion.
Symbiotic Miras are normally discovered as such when the WD companion enters
a powerful outburst that ionizes and lights-up by several magnitudes the
circumstellar material producing a rich emission line forest and a flat blue
continuum.

Symbiotic Miras with erupting WD companions show a slow, smooth and
monotonic decrease in brightness following outburst maximum. For example,
the extremely smooth decline of HM~Sge and V1016~Cyg over the last 20 years
has been characterized by $\Delta V \sim$0.05 and $\Delta V \sim$0.03 mag
yr$^{-1}$, respectively, thus a whole 1 mag for HM~Sge and 0.6 mag for
V1016~Cyg in 20 years.

On the contrary, K~3-9 has been very stable over the last half century:
Table~2 data, Palomar DSS-1 and DSS-2 blue plates ($B$=18.25 on JD=2434895,
and $B$=18.30 on JD=2445936) as well as mean values for 1999 from Henden and
Munari (2000) all cluster tightly around $B$=18.3, as shown in Figure~4.
The absence of any descending trend and the $\Delta B$=1.1 bright phase in 1959
are more reminiscent of classical symbiotic stars with moderate active
phases than of symbiotic Miras in outburst. K~3-9 bears differences with
symbiotic Miras in outburst also in the spectra: observations included in  
the multi-epoch spectrophotometric atlas of symbiotic stars by Munari and
Zwitter (2001) indicate a moderate ionization of a radiation-bounded
circumstellar enviroment, quite different from the situation in HM~Sge,
V1016 Cyg and other symbiotic Miras in outburst characterized by firework
displays of emission line spectra.

\subsection{V407 Cyg}

\begin{figure}
\centering
\includegraphics[width=8.8cm]{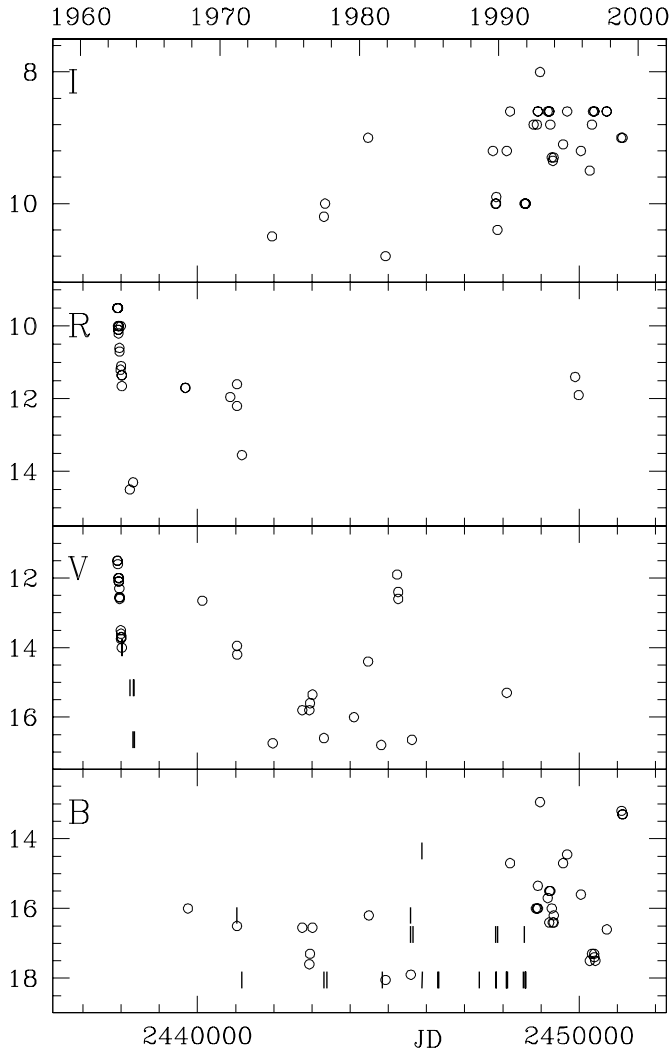}
      \caption{$B$, $V$, $R$ and $I$ lightcurves of V407 Cyg. Vertical bars 
               indicate {\sl fainter than}.} 
\label{V407CygA}
\end{figure}

Discovered as Nova Cyg 1936, V407~Cyg attracted interest when ($a$)
Meinunger (1966) reported that the main cause of the object's large variability
was the presence of a Mira with a 745 day pulsation period (too long for
normal field Miras), and ($b$) unpublished spectra by different authors
described a symbiotic star spectrum appearing from time to time (e.g.
Duerbeck 1987). Munari et al. (1990) investigated V407~Cyg lightcurve up to
late 80ies, confirming the presence of a Mira and discovering a large
modulation of the system mean brightness by $\Delta B \sim$3 mag. A possible
43 year periodicity was derived for the latter and interpreted as the system
orbital period in the framework of the Whitelock (1987) model of variable dust
obscuration. However, the lightcurve analyzed by Munari et al. spanned 51
years or just one possible orbital period. Thus the search for blue plates
obtained earlier than 1935 with astrographs able to reach $B$=16 would be
essential to investigate the secular behaviour of mean brightness and then
to confirm the possible 43 year periodicity.

Pulsation periods as long as that of V407~Cyg almost invariably pertain to
the central star of OH/IR sources, engulfed by extremely thick dust cocoons
that make these objects invisible at blue wavelengths. If the presence of
the accreting and outbursting white dwarf companion inhibits the
formation of all but an optically thin circumstellar dust shell (Kolotilov et al.
1998), then V407~Cyg would allow an unobstructed view of the central star of
an OH/IR source. The dust could partially form and survive only in the shadow
cone produced by the Mira itself, which blocks the destroying action of the
white dwarf hard radiation field. When, during the orbital motion, the Mira
passes at the inferior conjunction and the whole system is seen through the
dust cone, the optical pulsation curve is shifted toward fainter magnitudes.
In such a framework, the minimum of mean $B$ brightness in the early 70s
would correspond to passage of the Mira at the inferior conjunction.

A new outburst was discovered in 1994 (Munari et al. 1994), 58 years after
the discovery event in 1936. At that time the Mira's spectrum appeared
severely veiled by a hot continuum in the blue and strong lines of Hydrogen,
HeI, HeII, [OIII] erupted in emission, settling the
classification of V407~Cyg as a symbiotic star.

\begin{figure*}
\centering
\includegraphics[width=18cm]{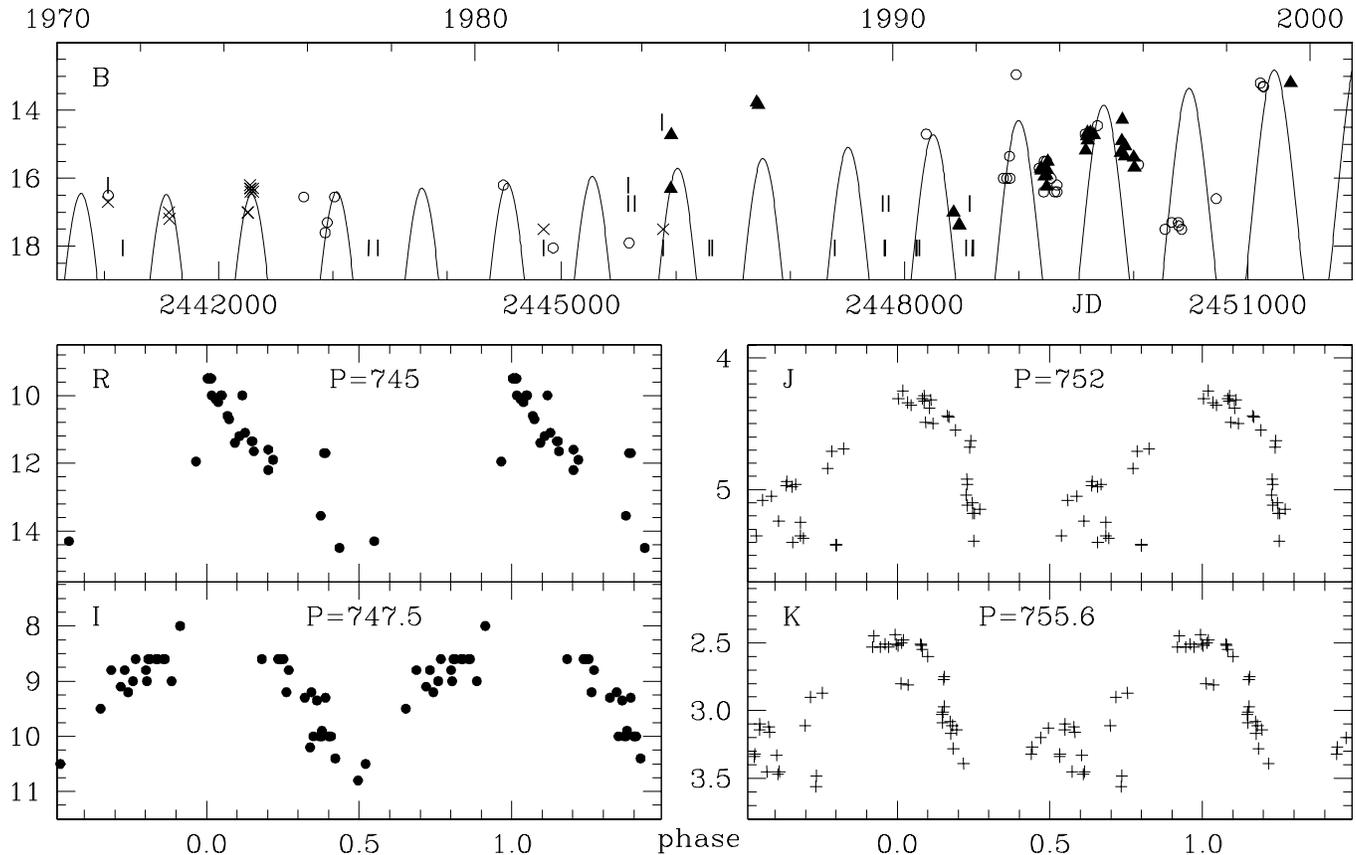}
       \caption{{\sl Upper panel}: $B$ lightcurve of V407 Cyg with
               overplotted a sinusoid phased according to Eq. (4) and
               modulated in mean brightness with a 43 year periodicity. 
               {\sl Lower panels}: phase plot of $R$ and $I$ data from
               Table~2 and $J$ and $K$ data from Kolotilov et al. (1998)
               according to Eq.s (5)--(8).}
\label{V407CygB}
\end{figure*}

The results of our searching the Asiago plate archive for V407~Cyg are
presented in Figure~5. Plates spanning a 40 year interval have been located.
The $B$ band data are plotted in a zoomed view in Figure~6 (top panel) and a
$\Delta mag =8$ amplitude sinusoid following the original Meinunger (1966)
ephemeris
\begin{equation}
Max(B)\ =\ 2429710\ +\ 745~~~\times E\\
\end{equation}
is overplotted for reference. A 43 year, $\Delta mag=2.1$ amplitude
modulation is superimposed to represent the dust obscuration phase that
peaked in 1973. The correspondence between observed points and the
sinusoidal approximation is not impressive, for at least two reasons: ($a$)
irregular variability by the accreting white dwarf may contribute
significantly to the overall system $B$ band brightness (apart from the 1936
and 1994 outbursts), and ($b$) the Mira in V407~Cyg does not follow regular
pulsation cycles characterized by similar and in-phase lightcurves (the
VSNET and VSOLJ amateur estimate databanks clearly show quite perturbed
lightcurves over the last fifteen years).

The complexity of the Mira pulsations in V407~Cyg is evident when its
behaviour at different wavelengths is compared (Figure~6). The lightcurve
in the $B$ band is only moderately periodic, while the periodicity looks instead
almost perfect in our $R$ and $I$ data. However, the shape and period of the
pulsation lightcurve change with the wavelengths. The $R$, $I$, $J$ and $K$
lightcurves in Figure~6 are computed according to the following ephemerids:
\begin{eqnarray}
Max(R)\ &=&\ 2429710\ +\ 745~~~\times E\\
Max(I)\ &=&\ 2446050\ +\ 747.5\times E\\
Max(J)\ &=&\ 2446100\ +\ 752~~~\times E\\
Max(K)\ &=&\ 2446150\ +\ 755.6\times E
\end{eqnarray}
where the original $B$-band ephemeris of Meinunger (1966) was found to
be the best performing one for the $R$ data and the Kolotilov et al. (1998)
ephemeris the best for the $J$ data. They are however not satisfactory for
$I$ and $K$ data, for which we determined new periodicities with
converging results by different techniques (Fourier and phase dispersion
minimization). The apparent increase of the pulsation period (by 1.4\%) and
shift of phase of maximum (by 13\%) going from $R$ to $K$ wavelengths seems
a robust one, as indicated by the poor results that are obtained if the $R$
data are phase plotted against Eq. (8) or the $K$ data against Eq. (5).

The differences highlighted by Eq.s (5)--(8), whose validity could vanish outside the
time-span covered by the data used to derive them, might be related to the highly
unusual evolutionary state of the Mira in V407~Cyg, resembling the
central star of OH/IR sources. Beating among phenomena characterized by
different periodicities and effective temperatures could induce the above
shifts in phase and period. Optical depth effects in the Mira's atmosphere
and circumstellar dust envelope could add to the overall picture: viewing
the system at different wavelengths means different transparencies of the
circumstellar dust and different effective photospheric levels in the
stellar atmosphere.

\subsection{V335~Vul, FG~Ser, Draco~C-1}

The remaining program stars are only briefly commented upon.

V335~Vul was discovered by Dahlmark (1993) to contain a carbon Mira of 342
day pulsation period. On the basis of marked spectral changes and the
emission line spectrum, Munari et al. (1999) suggested a possible symbiotic
nature for this object. Our $B$ and $V$ data in Table~2 reveal a variability
of significant amplitude, but the resulting lightcurve is disappointingly
restricted to a narrow range. This is caused by the short time span covered
by the plates (a couple of years), the long pulsation period (close to one
year), and the seasonal visibility.

The photometric and outburst history of the eclipsing symbiotic star FG~Ser
has already been reviewed in detail by Munari et al. (1992, 1995) and
Kurochkin (1993), and it will not be further commented upon here. Part of
the plates here used to derive the magnitude of FG~Ser were already
investigated by Munari et al., but they are measured again here both to place
them on the more accurate Henden \& Munari (2000) photometric sequence and
to link them with the more recent data reported in Table~2.

Similarly, Draco~C-1 plate material prior to 1991 has already been
considered by Munari (1991). It has been estimated again here against the
better Henden \& Munari (2000) comparison sequence, in order to place it on the
same scale as the data collected after 1991 and included in Table~2. The
star is {\sl very} close to the limiting magnitude on all plates, and this
affects both star detection and brightness estimate. A significant $\Delta
I$=0.7 mag variability is however evident, apparently not much correlated
with $B$ behavior. Apart from a $\Delta B$=1.3 mag bright phase in 1988,
Draco~C-1 remained quiescent around $B$=18.4 over 1987--1996, close to
the mean $B$=18.6 measured in 1999 by Henden \& Munari (2000) and the
$B$=18.43 measured in 1981 by Aaronson et al. (1982).

\begin{acknowledgements}
RJS wish to thank the hospitality of the Asiago observatory and the
Primorsko-Goranska County for financial support.
\end{acknowledgements}

\clearpage

\begin{table*} 
\centering
\caption[]{The {\sl UBV(RI)$_{\rm C}$} magnitudes of the program stars
estimated on the Asiago archive plates. The date is given in the
year/month/day format, the JD is heliocentric and
the magnitude is estimated in steps of 0.05 mag.}
\includegraphics[width=16cm]{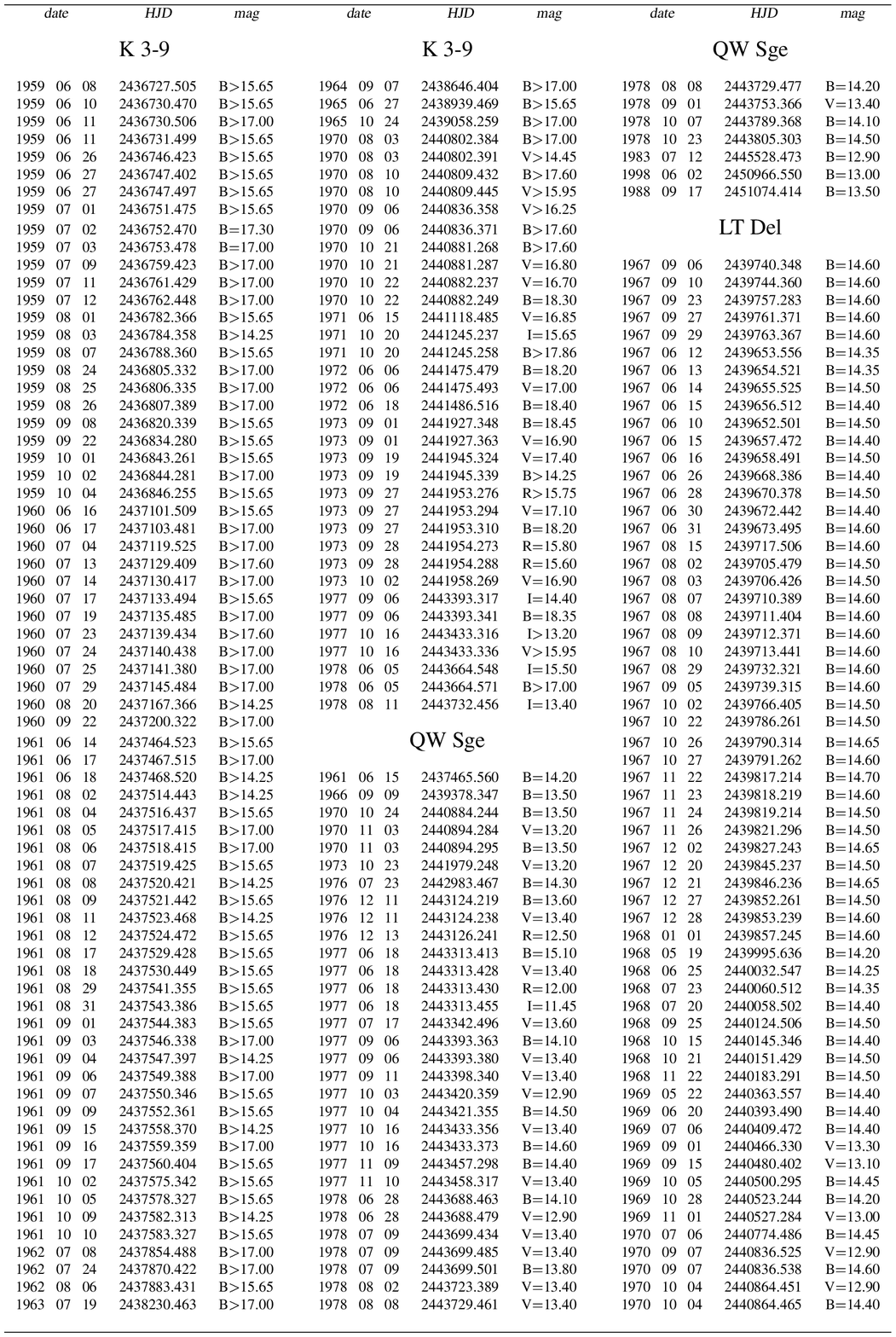}
\label{tab2a}
\end{table*}

\setcounter{table}{1}
\begin{table*} 
\centering
\caption[]{({\sl continues})}
\includegraphics[width=16cm]{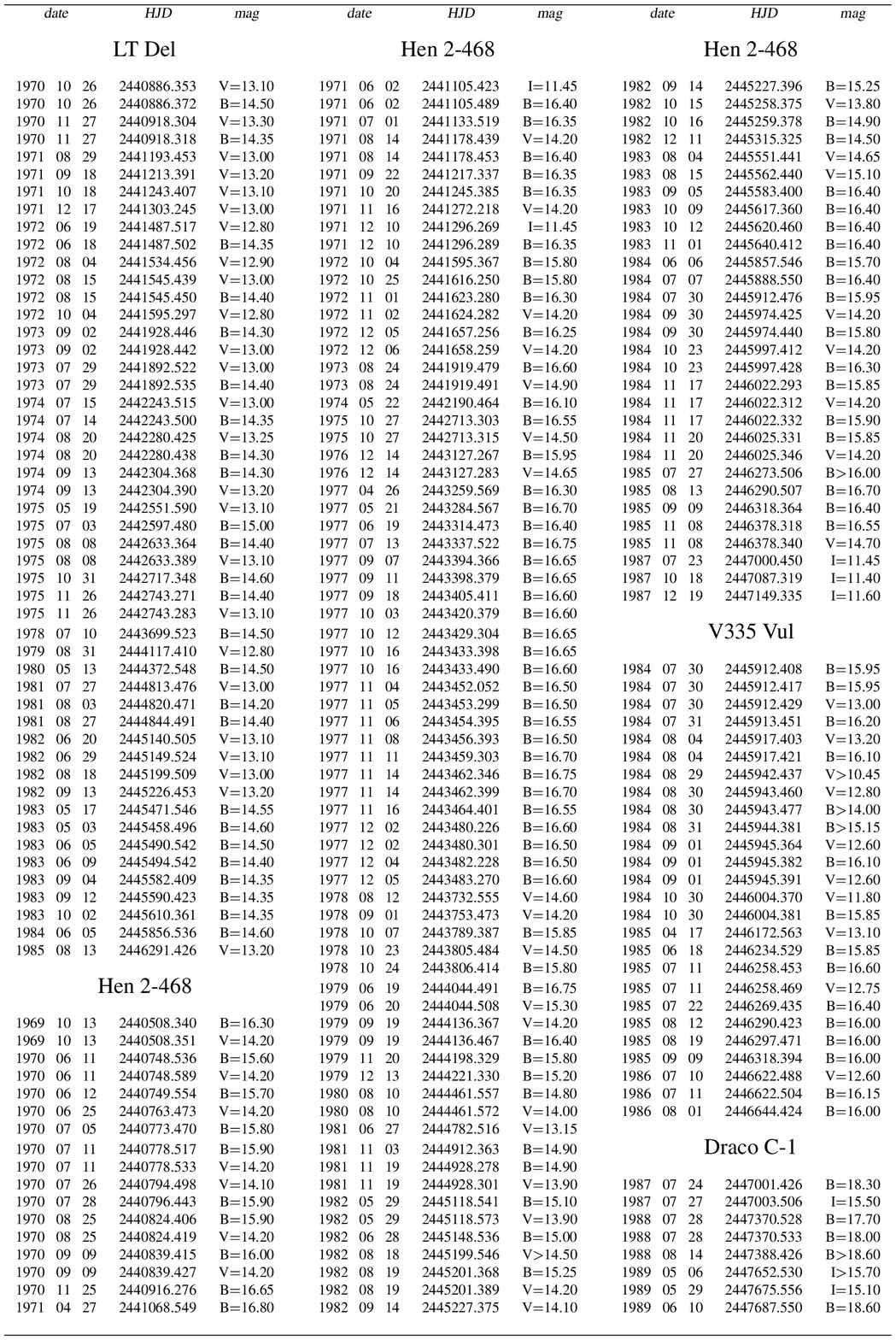}
\label{tab2b}
\end{table*}

\setcounter{table}{1}
\begin{table*} 
\centering
\caption[]{({\sl continues})}
\includegraphics[width=16cm]{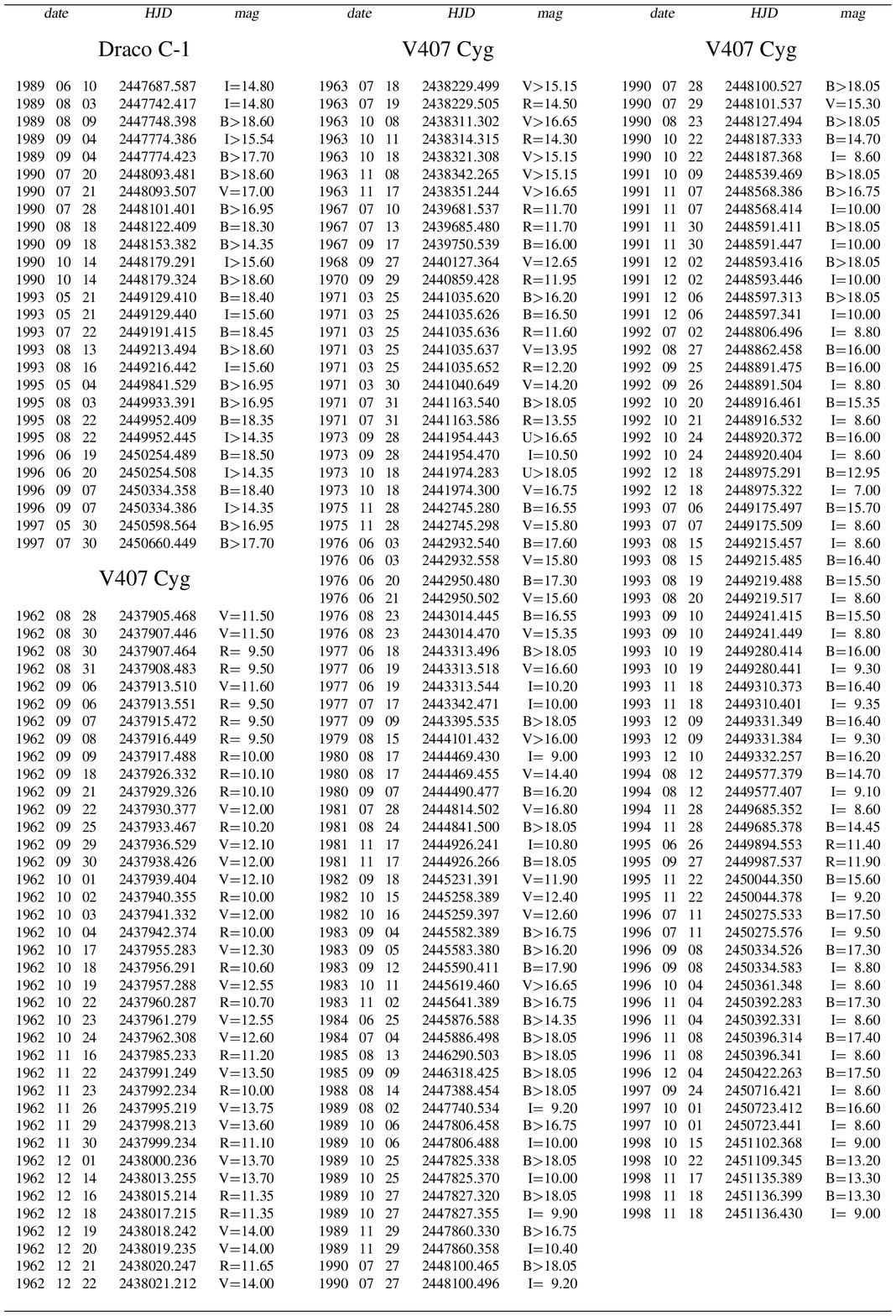}
\label{tab2c}
\end{table*}

\setcounter{table}{1}
\begin{table*} 
\caption[]{({\sl continues})}
\includegraphics[width=5cm]{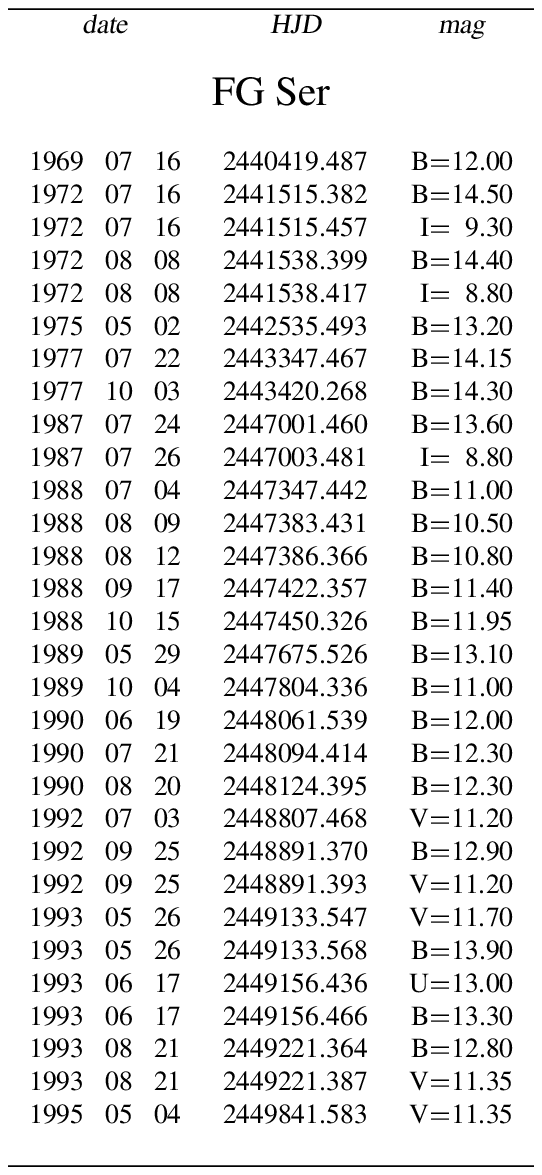}
\label{tab2d}
\end{table*}


\begin{thebibliography}{}

\bibitem{} Allen D.A., 1974, MNRAS 168, 1
\bibitem{} Allen D.A., 1984, Proc.A.S.A. 5, 369
\bibitem{} Aaronson M., Liebert, J., Stocke, J., 1982, Ap.J. 254, 507
\bibitem{} Arkhipova V.P., Noskova, R.I., 1988 SvA Lett 14, 188
\bibitem{} Arkhipova V.P., Ikonnikova, N.P., Noskova, R.I. 1995, PAZh 21, 379
\bibitem{} Carrasco L., Costero R., Serrano P.G. 1983, in {\it Planetary Nebulae},
    Proc. of IAU Symp. 103, D.R.Flower ed., pag. 548
\bibitem{} Dahlmark L., 1993, IBVS 3855
\bibitem{} Duerbeck, H.W. 1987, A Reference Catalogue and Atals of Galactic Novae, Reidel, Dordrecht
\bibitem{} Henden A.A., Munari U., 2000, A\&AS 143, 343
\bibitem{} Henden A.A., Munari U., 2001, A\&A 372, 145
\bibitem{} Henden A.A., Munari U., 2002, A\&A to be submitted
\bibitem{} Henize K.G. 1967, ApJS 14, 125
\bibitem{} Ivison R.J., Seaquist E.R. 1995, MNRAS 272, 878
\bibitem{} Kenyon 1986, The Symbiotic Stars, Cambridge University Press
\bibitem{} Kolotilov E.A., Munari U., Popova A.A., Tatarnikov A.M., Shenavrin V.I., Yudin B.F. 1998, Astron. Lett. 24, 451
\bibitem{} Kurochkin N.E. 1993, Astron.Astrophys.Transactions 3, 295
\bibitem{} Margoni R., Stagni R. 1984, A\&AS 56, 87
\bibitem{} Meinunger L., 1966, Mitt.Veranderl.Sterne 3, 111
\bibitem{} Mikolajewska J. 1996, Physical Processes in Symbiotic Binaries and Related Objects,
           editor, Polish Academy of Sciences, Warsaw
\bibitem{} Munari U., 1991, A\&A 251, 103
\bibitem{} Munari U., Buson L.M.  1991, A\&A 249, 141
\bibitem{} Munari U., Buson L.M.  1992, A\&A 255, 158
\bibitem{} Munari U., Zwitter T.  2002, A\&A in press
\bibitem{} Munari U, Bragaglia A., Guarnieri M.D., Sostero G., Lepardo A.,
           Yudin B.F., 1994, IAU Circ 6049
\bibitem{} Munari U., Jurdana-\v{S}epi\'c R., Moro D. 2001, A\&A 370, 503 (Paper I)
\bibitem{} Munari U., Margoni R., Stagni R.  1990, MNRAS 242, 653
\bibitem{} Munari U., Tomov T., Rejkuba M., 1999, Inf.Bull.Var.Stars 4668
\bibitem{} Munari U., Rejkuba M., Mattei J., Hazen M., Luthardt R., Yudin B.F. 1997,
           A\&A 323, 113
\bibitem{} Munari U., Yudin B.F., Taranova O.G., Massone G., Marang F., Roberts G.,
           Winkler H., Whitelock P.A.  1992a, A\&AS 93, 383
\bibitem{} Munari U., Yudin B.F., Kolotilov E., Gilmore A.  1995, AJ 109, 1740
\bibitem{} Munari U., Whitelock P.A., Gilmore A.C.,  Blanco C., Massone G.,
           Schmeer P.K.  1992b, AJ 104, 262
\bibitem{} Passuello R., Saccavino S., Munari U. 1994, IAU Circ 6065
\bibitem{} Whitelock P.A., 1987 PASP 99, 573
\end{thebibliography}
\end{document}